\documentclass[12pt, a4paper]{article}
\input epsf
\newcommand{\sect}[1]{ \section{#1} }

\newcommand{\ve}{\left( \begin{array}{r}}
\newcommand{\ev}{\end{array} \right)}
\newcommand{\ar}{\left( \begin{array}{rr}}
\newcommand{\ra}{\end{array} \right)}
\newcommand{\arr}{\left( \begin{array}{rrrr}}
\newcommand{\arrr}{\left( \begin{array}{rrrrrr}}
\newcommand{\eqr}{\begin{eqnarray}}
\newcommand{\rqe}{\end{eqnarray}}
\newcommand{\eq}{\begin{equation}}
\newcommand{\qe}{\end{equation}}
\newcommand{\half}{\frac{1}{2}}

\newcommand{\eps}{\epsilon}

\newcommand\preprint[1]{\vspace{-1in}\vtop{\null\hfill
\parbox[t]{1.6in}{\small\sc #1\\\null}}
\vskip .5in\bigskip\normalfont}
\def\KK{{\rm I\kern -.2em  K}}
\def\NN{{\rm I\kern -.16em N}}
\def\RR{{\rm I\kern -.2em  R}}

\def\ZZZ{{\small{\rm Z}\kern -.5em Z}}
\def\QQ{{\rm \kern .25em
             \vrule height1.4ex depth-.12ex width.06em\kern-.31em Q}}
\def\CC{{\rm \kern .25em
             \vrule height1.4ex depth-.12ex width.06em\kern-.31em C}}

\usepackage{epsfig}

\title{\preprint{MIT-CTP-3262\\
hep-th/0207129}Supersymmetric webs of D3/D5-branes in supergravity}
\author{ Ansar Fayyazuddin\footnote{email: ansar@lns.mit.edu}}

\begin{document}
\maketitle
\begin{center}
{\em
\vspace{-0.5cm}
Center for Theoretical Physics\\
Massachusetts Institute of Technology\\
Cambridge, MA 02139\\
\vspace{0.5cm} 
and\\
\vspace{0.5cm}
Department of Physics\\
Stockholm University\\
Box 6730\\
S-113 85 Stockholm\\
Sweden 
}

\end{center}

\vspace{0.5cm}

\begin{abstract}
We study webs of D3- and D5-branes in type IIB supergravity.  
These webs preserve at least
8 supercharges.  By solving the Killing spinor equations we determine
the form of supergravity solutions for the system. 
We then turn to the sub-class of the intersecting D3/D5 brane system and 
elucidate some of its features.
 
\end{abstract}

\vspace{-18.5cm}

\thispagestyle{empty}

\newpage

\setcounter{page}{1}

\sect{Introduction}
In this paper we examine webs of D3-branes and D5-branes from the point 
of view of supergravity.  There are at least three motivations for 
studying this system.
The first motivation is to remedy our lack of understanding of localized 
intersections of branes \cite{gaunt}
and more generally of branes ending on branes \cite{Strominger:1995ac} in
supergravity.  Examples of such systems may help in 
unravelling some of their features.
The second originates in Hanany and Witten's \cite{hw} construction 
of 3-dimensional N=4 
gauge theories using webs of NS5-branes and D3-branes, the S-dual of which are
the webs studied here.  A Maldacena type limit \cite{malda}
of the supergravity
solution of such brane configurations should yield supergravity duals of 
these theories.  These supergravity duals may help elucidate 
non-perturbative aspects of the gauge theories.  The third motivation comes
from the recent work of Karch and Randall \cite{kr} on constructing 
localized gravity models on branes and more generally branes in 
asymptotically AdS geometries. In these models one views the D5-branes as 
domain walls in the near-horizon
AdS$_5$ geometry of the D3-branes.  Such systems are of interest both because
of their cosmological implications and also for their intrinsic interest as
string theory backgrounds. 

The main results of this paper are as follows.  We develop a supersymmetric 
ansatz for the fields in type IIB supergravity capable of accomodating 
webs of D3/D5-branes.  The metric and form fields have certain interesting
features which we elucidate.  Using the ansatz we study in closer detail the
case of the intersecting
D3/D5-brane system.  We present a simple but non-linear set of equations 
for this sub-case.  We present solutions to this system of equations in the 
smeared D5-brane approximation in certain asymptotic regimes.

The structure of the paper is as follows.  In section 2 we describe the brane
set-up and the supersymmetry preservation conditions in the probe 
approximation.  In section 3 we propose an ansatz for the metric and
examine the Killing spinor
equations in type IIB supergravity with appropriate fields turned on and 
determine a class of supersymmetric field configurations.  Then in section 4 
we restrict ourselves to the sub-class of intersecting branes and remark on
the structure of supergravity solutions for these configurations.  We set up
the Bianchi and source equations and solve them in the smeared approximation
in certain asymptotic regimes.  In section 5 we end with some concluding 
remarks.

\section{Webs of D3/D5-branes}
Consider a web of D3-branes and D5-branes where the D3-branes are along the
$x^1,x^2,x^6$ directions while the D5-branes are along the 
$x^1,x^2,x^3,x^4,x^5$
directions.  The D5-branes are located at $x^7=x^8=x^9=0$ and separated in
the $x^6$ direction.  The D3-branes, on the other hand, are located at
$x^7=x^8=x^9=0$ but can be seperated in the $x^3, x^4, x^5$ directions.
The D3-branes end on the D5-branes and stretch in the $x^6$ direction between 
adjacent D5-branes.  When the D3-branes are aligned to form infinite D3-branes
they can be lifted off the D5-branes.  This important sub-case of infinite
D3-branes will be referred to as the intersecting branes case.  

The S-dual of the above set-up is known as the Hanany-Witten 
system\cite{hw}, it
was used by the authors to construct $N=4$ Yang-Mills in string theory.  
We review some features below.

The S-dual of the D3/D5-brane system is the system described above but with
the D5-branes replaced by NS5-branes.  The 
D3-branes have four dimensional worldvolume theories truncated in the 
$x^6$ direction by the NS5-branes.  Thus macroscopically, or at low energies, 
the D3-brane worldvolume theories are effectively a product
of three dimensional gauge
theories with gauge groups $SU(n_i)$, where $n_i$ is the number
of D3-branes stretched between the ith and (i+1)th NS5-branes.  
The coupling constants of the gauge theories are
$1/g_{YM,i}^2 = L_i/g_s$ where $L_i$ is the coordinate distance in $x^6$ 
between the ith and (i+1)th NS5-brane and $g_s$ is the string coupling 
constant. 
The $L_i$ specify UV cut-offs in real space 
(since probing distances smaller
than the cut-off will allow one to probe the 4-th dimension).  The limit in
which we take the $L_i \rightarrow 0$, is the limit in which the Yang-Mills
coupling becomes infinitely strong.  This limit corresponds to an infrared 
limit of the quatnum field theory.

The positions of the D3-branes in the $x^3, x^4, x^5$ directions parameterize
the Coulomb branch of the gauge theory living on the D3-branes, while 
also acting as bare masses
of the fundamental matter multiplets of the theories living on adjacent
intervals.  Finally, the positions of the NS5-branes in the overall transverse
direction (fixed to be at the origin in everything that follows) 
parameterize Fayet-Illiopoulis terms.

In the remainder of the paper we will be studying the S-dual of the
D3/NS5-brane system described above.  
S-duality in the field theory acts by dualizing the gauge-field 
into a periodic scalar and inverting the Yang-Mills coupling constant.  
On the supergravity side S-duality replaces the NS5-branes with D5-branes and
inverts the string coupling.  

One can view the above brane set-up in a probe approximation in which the
influence of the branes on the bulk geometry is neglected.  This is a useful 
approximation when trying to determine the amount of preserved supersymmetry.  
Essentially one introduces open strings with appropriate
boundary conditions for the above brane configuration in flat Minkowski space
and determines the amount of supersymmetry that is preserved \cite{pol1,pol2}.
For Dp-branes
the supersymmetry preservation condition is \cite{pol1,pol2}:
\eq
\epsilon_L = {\hat \Gamma}_{012..p}\epsilon_R
\qe
where the ${\hat \Gamma}$ are flat space $\Gamma$-matrices, and $\epsilon_L,
\epsilon_R$ are worldsheet left- and right-handed space-time spinors acting
as supersymmetry variation parameters.
In the case at hand the amount of space-time supersymmetry preserved 
was determined
in \cite{hw}, we repeat the calculation here for completeness.  The 
variation parameters satisfy:
\eqr
\epsilon_L& = &{\hat \Gamma}_{0126}\epsilon_R \nonumber \\
\epsilon_L& = &{\hat \Gamma}_{012345}\epsilon_R.
\rqe 
The spinors $\epsilon_L, \epsilon_R$ are Majorana-Weyl fermions satisfying:
\eqr
\epsilon_L& = &{\hat \Gamma}_{0123456789}\epsilon_L \nonumber \\
\epsilon_R& = &{\hat \Gamma}_{0123456789}\epsilon_R.
\rqe
The above constraints are satisfied by the subspace of spinors which can
be parameterized by $8$ real parameters.  This corresponds to $N=4$
supersymmetry in $d=3$.
It is convenient to combine the spinors into a single complex spinor as
follows:
\eq
\epsilon = \epsilon_R + i\epsilon_L.
\qe
Then the above conditions imply:
\eqr
-i\epsilon & = &{\hat \Gamma}_{0126}\epsilon \nonumber \\
-\epsilon^{*}& = &{\hat \Gamma}_{3456}\epsilon    \\
i\epsilon^*& = &{\hat \Gamma}_{6789}\epsilon. \nonumber \\
\label{probe}
\rqe
$*$ denotes complex conjugation.

We will assume that the constraints on the supersymmetry variation
parameters obtained in the probe approximation continue to apply in 
the full solution.  These constraints will be used in the next section to
obtain supersymmetric solutions for the D3/D5 system following broadly the
methods of \cite{fs} (see also \cite{Rajaraman:2000ws} for the case
of strings ending on D-branes).

\section{Supersymmetric ansatz for the supergravity solution}
One approach to solving supergravity equations in situations where some
fraction of the supersymmetry is preserved is to solve the Killing spinor
equations rather than the full set of non-linear Einstein equations coupled
to matter.  These Bogomolny type equations tend to be considerably simpler.
I will take this approach in what follows.

Type IIB supergravity has the following massless field content.  There are 
in the
NS-NS sector, the dilaton, metric, and NS-NS 2-form.  In the R-R sector there
is a scalar (also known as the axion), a 2-form, and
a 4-form with self-dual field strength. We will use the notation of 
\cite{schwarz}
There are two Killing spinor equations in type IIB supergravity: one for the 
dilatino and one for the gravitino:
\eqr
\delta\lambda &=&\frac{i}{\kappa}\Gamma^{M}P_M\eps^{*}-\frac{i}{24}
\Gamma^{MNP}G_{MNP}\eps =0\nonumber\\
\delta\psi_M &=&\frac{1}{\kappa}(D_M-\frac{i}{2}Q_M)\eps + \frac{i}{480}\Gamma^{M_1...M_5}F_{M_1...M_5}\Gamma_M\eps =0.
\rqe

The strategy in the remainder of this section will be to develop a
sufficiently general ansatz for the supergravity fields to accomodate 
the D3/D5-brane system.  To proceed we first set some notation.  We will denote
$x^6$ by $y$; indices in the beginning of the Greek alphabet (i.e. $\alpha, 
\beta, \gamma$ etc) will take values in $3,4,5$ (directions transverse to the
D3-brane but along the D5-brane); Greek indices in the middle of the alphabet
(i.e. $\mu, \nu, \rho$ etc) will take values in $0,1,2$ (these are directions
shared by the D3- and D5-branes); finally Latin indices in the middle of
the alphabet (i.e. $i,j,k$ etc) will take values in $7,8,9$ (these directions
are transverse to all branes).  Hatted indices are flat indices, thus:
$G_{{\hat m}{\hat n}{\hat p}} \equiv G_{MNP}e_{\hat m}^{M}e_{\hat n}^{N}
e_{\hat p}^{P}$.

We take the Einstein frame metric to be of the following general form:
\eq
ds_E^2 = H_1^2\eta_{\mu\nu}dx^\mu dx^\nu + g_{\alpha\beta}dx^\alpha dx^\beta
+ D^2(dy + A_\alpha dx^\alpha)^2 + H_2^2\delta_{ij}dx^i dx^j.
\qe
In the above ansatz we allow for general off-diagonal terms between
coordinates in the $3,4,5,6$ directions.  We have collected the off-diagonal
terms involving $y$ and the $3,4,5$ direction as a vector field in the
$3,4,5$ directions.  In the overall transverse directions
we have a diagonal metric.  For the solution to be Lorentz invariant in the
$0,1,2$ directions we take $H_1, H_2, D$ to be independent of these 
directions. We will set the Ramond-Ramond scalar to zero, thus
$\tau = i\tau_2$.  In addition we will assume that the only non-zero 
$G_{{\hat m}{\hat n}{\hat p}}$ are ones with the following index structure:
$G_{{\hat 3}{\hat 4}{\hat 5}}, G_{{\hat 7}{\hat 8}{\hat 9}}, 
G_{{\hat y}{\hat\alpha}{\hat\beta}}, G_{{\hat y}{\hat i}{\hat j}}$.  
Similarly, for the five-form we will assume that the only non-zero components
are those with index structure $F_{{\hat 0}{\hat 1}{\hat 2}{\hat m}{\hat n}}$,
where $m,n$ take any value other than $0,1,2$. Self-duality of the five-form
will, of course, require other components to be non-zero as well.

Inserting the above ansatz into the Killing spinor equations and using the
relations (\ref{probe}) yields a set of relations between the
supergravity fields and $H_1, H_2, D$ and $A_\alpha$ appearing in
the metric.  Here we
present the solution to these constraints.  It is convenient to define the
combinations $K\equiv H_1^{-2}H_2^2$ and $H\equiv H_1^4H_2^4$:
\eqr
D^2 & = &H^{5/4}K^{-1/2} \nonumber \\
\tau_2& = &g_s^{-1}H^{1/2} \\
\partial_iA_\alpha & = &0 \nonumber \\
0 & = &\partial_\alpha A_\beta - \partial_\beta A_\alpha - 
A_\alpha\partial_y A_\beta + A_\beta\partial_y A_\alpha  \nonumber \\
g_{\alpha\beta}& = &H^{-3/4}K^{1/2}\delta_{\alpha\beta}\nonumber \\
\partial_y(A_{\beta} H) & = & \partial_\beta H
\rqe
In addition one finds expressions for the self-dual 5-form, and the SL(2,R)
invariant 3-form:
\eqr
F_5 & = &\frac{1}{4\kappa}H(\partial_\beta K^{-1}-A_\beta\partial_yK^{-1})
dx^0\wedge dx^1\wedge dx^2\wedge dy\wedge dx^\beta \nonumber \\
&+ &\frac{1}{4\kappa}\partial_j(HK^{-1})dx^0\wedge dx^1\wedge dx^2\wedge 
dy\wedge dx^j \nonumber \\
& - &\frac{1}{8\kappa}\epsilon_{\alpha\beta\delta}(
2H^2K A_\alpha\partial_\beta K^{-1}dx^\delta\wedge dy -
\{H^2K(A_\alpha A\gamma\partial_\gamma K^{-1} \nonumber \\
& - & A_\gamma A_\gamma\partial_\alpha K^{-1}) 
+ \partial_\alpha K - A_\alpha\partial_yK\}dx^\beta\wedge dx^\delta )
\wedge dx^7\wedge dx^8\wedge dx^9   \\
& + & \frac{1}{8\kappa}\epsilon_{ijk}\partial_i(HK^{-1})
( (K^2H^{-2} + KA_\gamma A_\gamma )
dx^3\wedge dx^4\wedge dx^5 \nonumber \\
& + &\half KA_\alpha\eps_{\alpha\beta\gamma} 
dy\wedge dx^\beta\wedge dx^\gamma )\wedge dx^j\wedge dx^k, \nonumber
\rqe 
and
\eqr
G & = &\frac{1}{\kappa}(-2H^{-9/4}K^{3/2}\partial_y(H^{1/2}K^{-1/2})+ 
H^{1/4}A_\gamma\partial_yA_\gamma)dx^3\wedge dx^4\wedge dx^5 \nonumber \\
& + &\frac{1}{2\kappa}H^{1/4}\partial_yA_\delta dy\wedge dx^\alpha\wedge 
dx^\beta \\
& + &i(-\frac{2}{3\kappa}\eps_{ijk}\partial_kH^{3/4}dy\wedge x^i\wedge dx^j
+ \frac{2}{\kappa}H^{-1/4}K^{1/2}\partial_yK^{1/2}dx^7\wedge dx^8\wedge dx^9).
\nonumber
\rqe

From these expressions we can extract the string frame metric and the
3-form field strengths in the R-R and NS-NS sectors.  They are as follows:
\eqr
ds^2 & = &K^{-1/2}\eta_{\mu\nu}dx^\mu dx^\nu + HK^{-1/2}(dy + A_\alpha 
dx^\alpha)^2 \nonumber \\
& + &H^{-1}K^{1/2}\delta_{\alpha\beta}dx^\alpha dx^\beta + 
K^{1/2}\delta_{ij}dx^i dx^j, \nonumber \\
D_{3} & = & \frac{1}{\kappa}(-\frac{1}{2}\eps_{ijk}\partial_kH dy\wedge dx^i
\wedge dx^j + \partial_yK dx^7\wedge dx^8\wedge dx^9),\nonumber \\
H_{3} & = & \frac{1}{\kappa}(-2H^{-5/2}K^{3/2}\partial_y(H^{1/2}K^{-1/2})
+ A_\gamma\partial_y A_\gamma)dx^3\wedge dx^4\wedge dx^5\nonumber \\
& + & 
\frac{1}{2}\partial_yA_\delta
\eps_{\alpha\beta\delta} dy\wedge dx^\alpha\wedge dx^\beta).
\rqe

Some comments are in order here.  The off-diagonal components of the metric
are completely controlled by the one-form $A$.  When $A$ vanishes
the metric is completely diagonal.  Although the form of the metric and the
constraint that $A$ be independent of the $x^i$ coordinates
suggests an interpretation of $A$ as a gauge-field, the metric in
general will not have a Killing symmetry corresponding to translations in
the $y$ direction.  This is because
in solutions in which the D5-branes are localized in $y$, the
putative isometry is explicitly broken.  However, below we will 
consider the case
in which the D5-branes are smeared in $y$ producing a
Killing symmetry which allows for the interpretation of
$A$ as a gauge field.  We also note that the NS-NS
3-form field-strength is non-zero, in general, despite the absence
of NS-NS 5-branes, this is a feature our solution shares with
the Klebanov-Strassler solution and its generalizations.

The above solutions to the Killing spinor equations guarantee solutions
to the Einstein equations. However they have to be supplemented by source
equations and Bianchi identities for the 5-form and 3-form field strengths.  
These equations are:
\eqr
dF_5 & = & \frac{\kappa}{4}D_{3}\wedge H_3, \nonumber \\
dH_3 & = &0,\nonumber \\
dD_3 & = & 0, \\
d*_E(\tau_2H_3) & = & 4\kappa F_5\wedge D_3,  \nonumber \\
d*_E(\tau_2^{-1}D_3) & = & -4\kappa F_5\wedge H_3,  \nonumber\label{eom}
\rqe
where the $*_E$ is the Hodge star with respect to the Einstein metric.  

\subsection{Some checks}
To gain some confidence in our solution we perform some elementary checks.
We will simply try to reproduce the cases in which one has either D3-branes
or D5-branes.  These are
subcases of our more general set-up in which one of the R-R charges vanishes.
The subcase with only D3-branes consists of the following solution of
all the constraints:
\eqr
H &=& 1\nonumber \\
K &=& (1) + \frac{R_0^4}{(\sum_{\alpha}x_\alpha^2 +
\sum_{i}x_i^2)^2} \nonumber \\
A_\alpha &=& 0.
\rqe  
Similarly, the case with only D5-branes is reproduced by the following:
\eqr
H & = & K = (1) +\frac{R_0^2}{\sum_{i}x_i^2 + y^2}\nonumber \\
A &=& 0.
\rqe
The parentheses around the number $1$ in the above equations
indicates that they are present in the assymptotically flat cases and absent
in the near-horizon ones.  Since the D5-brane solution is correctly reproduced
an S-duality transformation automatically gives the correct NS5-brane solution
as well.

Having convinced ourselves that the known subcases are correctly 
reproduced, we can proceed to study new cases which have not hitherto
been studied in the literature.  In the following we will exclusively study
the case of intersecting branes, but we emphasize that we believe that
the ansatz is general enough to accomodate more general D3/D5-brane webs.

\section{Intersecting branes}
In this section we study the important subcase of a stack of coincident 
D3-branes intersecting one or more D5-branes.  We will set the problem up
for localized intersections, but will be forced to shift our attention to 
the more tractable problem of D5-branes smeared in the $y$ direction.

Consider the case, then, of coinciding $N$ D3-branes located at
$x^3= x^4=x^5=x^7=x^8=x^9=0$.  This is a special point in
the moduli space of possible configurations.  One can separate the
D3-branes along $x^3, x^4, x^5$ while continuing to intersect the D5-branes, 
in addition one can break them along the D5-branes.  
However, we will only focus on this special point in moduli space where
there is an
enhanced SO(3) symmetry which rotates the $x^\alpha$ into each other. When 
describing semi-infinite or finite D3-branes ending on D5-branes, the 
SO(3) symmetry is broken due to the bending of the D5-branes caused by
the tension of the D3-branes.  In the special case of infinite D3-brane
intersections the pulling of the D5-branes on one side is completely canceled
by the D3-branes on the other side, leaving the SO(3) symmetry intact.

Before proceeding to setting up the problem for intersecting branes, we set
some notation.  We define the usual spherical coordinates in the
three dimensional spaces spanned by the $x^\alpha$ and $x^i$.  In the
former we denote the radial coordinate by $r$ and the metric on
the unit sphere by $d\Omega_1^2$, while in the latter we denote the radial 
direction by $\rho$ and the metric on the unit sphere by $d\Omega_2^2$.  We
also define the volume form on the two unit spheres by $\omega^1$ and 
$\omega^2$.  Using these definitions,
the most general string-frame metric consistent with supersymmetry and
the SO(3)$\times$SO(3) symmetry is:
\eqr
ds^2 & = &K^{-1/2}\eta_{\mu\nu}dx^\mu dx^\nu + HK^{-1/2}(dy + A_r dr)^2 \\
& + &H^{-1}K^{1/2}(dr^2 + r^2d\Omega_1^2) + 
K^{1/2}(d\rho^2+\rho^2d\Omega_2^2), \nonumber \\
\rqe  
provided that $H=H(y,r,\rho)$ and $K=K(y,r,\rho)$.  That is, $H$ and $K$ do not
depend on the coordinates on the sphere.  Notice also that the 
SO(3)$\times$SO(3) symmetry requires that only the radial component of
the one-form $A$ be non-zero.  Had there been a Killing vector 
$\partial /\partial y$, the one-form could be set to zero.  However, as
stated above this is not the case for D5-branes localized in $y$.  

The form-fields can be written as follows:
\eqr
F_5 & = &\frac{1}{4\kappa}H(\partial_rK^{-1}-A_r\partial_yK^{-1})
dx^0\wedge dx^1\wedge dx^2\wedge dy\wedge dr \nonumber \\
&+ &\frac{1}{4\kappa}\partial_\rho (HK^{-1})dx^0\wedge dx^1\wedge dx^2\wedge 
dy\wedge d\rho \nonumber \\
& + &\frac{1}{4\kappa}r^2\rho^2(\partial_r K - A_r\partial_yK)
\omega^1\wedge\omega^2\wedge d\rho   \\
& + & \frac{1}{4\kappa}r^2\rho^2\partial_\rho (HK^{-1})
( (K^2H^{-2} + KA_r^2 )dr\wedge\omega^1 + 
KA_r dy\wedge\omega^1)\wedge\omega^2, \nonumber
\rqe
and 
\eqr
D_{3} & = & \frac{1}{\kappa}\rho^2
(-\partial_\rho H dy\wedge\omega^2
+ \partial_yK d\rho\wedge\omega^2),\nonumber \\
H_{3} & = & \frac{1}{\kappa}r^2((-2H^{-5/2}K^{3/2}\partial_y(H^{1/2}K^{-1/2})
+ A_\gamma\partial_y A_\gamma)dr\wedge\omega^1 \\
 &+& \partial_yA_r dy\wedge\omega^1).\nonumber
\rqe
These expressions are somewhat simpler than the general case.  The constraint
on the one-form $A$ can be expressed as:
\eqr
\partial_y(HA_r) & = &\partial_rH \nonumber \\
\partial_\rho A_r & = &0.
\rqe 
The equations of motion (\ref{eom}) are also simpler than the general case.
The first equation in (\ref{eom}) implies the following relations:
\eqr
0 &=& \partial_\rho\partial_y(A_rHK^{-1}) \nonumber \\
0 &=& \frac{1}{r^2}\partial_r\{ r^2(\partial_rK - A_r\partial_yK)\}
+\frac{1}{\rho^2}\partial_\rho\{\rho^2(\partial_\rho (H^{-1}K)-
A_r^2K\partial_\rho(HK^{-1}))\} 
\nonumber \\
&+& \partial_yK\{H^{-1}\partial_y(KH^{-1})+A_r\partial_yA_r\} \nonumber \\
0 &=& \partial_y\partial_rK- A_r\partial_y^2K -
\frac{1}{\rho^2}\partial_\rho\{\rho^2A_rK\partial_\rho (HK^{-1})\} \\
0 &=& \partial_y\{\partial_\rho(H^{-1}K)-\partial_\rho(HK^{-1})KA_r^2\}
+\frac{1}{r^2}\partial_r\{r^2A_rK\partial_\rho(HK^{-1})\} \nonumber \\
&-&\partial_\rho H\{H^{-1}\partial_y(KH^{-1})+A_r\partial_yA_r\}
\rqe
The second equation in the above set of equations is the source equation
for the D3-branes and one should have in mind a delta function
on the right hand side determining the location of the D3-branes.
Similarly, the Bianchi identities for the R-R and NS-NS three-forms impose the
following identities:
\eqr
0 &=& \partial_\rho(\rho^2\partial_\rho H) + \rho^2\partial_y^2K \nonumber \\
0 &=& \partial_\rho\partial_rH \nonumber \\
0 &=& \partial_r\partial_yK \\
0 &=& \partial_\rho\{H^{-1}\partial_y(KH^{-1})\} \nonumber \\
0 &=& \partial_y\{H^{-1}\partial_y(KH^{-1}) + A_r\partial_yA_r -
\frac{1}{r^2}\partial_r(r^2A_r)\} \nonumber \label{bianchi}
\rqe
In the above equations the first equation should really be thought of as
a source equation for the D5-branes.  The remaining are identities.
The last two equation in (\ref{eom}) give the following two equations:
\eqr
0&=&\partial_r(H\partial_y\ln (KH^{-1})) + \partial_y\{H(\partial_yA_r-
A_r\partial_y\ln (KH^{-1}))\}\\
&-& H(\partial_rK^{-1}-A_r\partial_yK^{-1})\partial_yK \label{H}\\
0 &=&
-\frac{1}{r^2}\partial_r(r^2K^{-1}A_r\partial_\rho H) - \partial_\rho
(H^{-1}K^{-1}\partial_y K)+\partial_y\{(A_r^2K^{-1}+H^{-2})
\partial_\rho H\}\nonumber \\
&+&\{H^{-1}\partial_y(KH^{-1})+A_r
\partial_yA_r\}\partial_\rho (HK^{-1})\nonumber 
\rqe
The last of these two equations is the equation of motion for the R-R 3-form
field strength.  Since we have D5-branes in the problem there are magnetic 
sources for the 3-form.  As is usual in such situations one should view this
equation as a Bianchi identity rather than as an equation of motion.
 
The above complex of equations along with appropriate boundary conditions 
should determine solutions of localized D3/D5-brane intersections.  If one
simplifies the above equations one finds that (\ref{H}) is satisfied when
\eq
0 = A_rHK^{-1}\partial_y^2K.
\qe
Thus either $A_r=0$ or $\partial_y^2K=0$. The second option yields delocalized
D5-branes as one can see from the source equation for the D5-brane (the first
equation in (\ref{bianchi})).  Thus $A_r=0$ is required.

The final set of equations that need to be satisfied for intersecting 
D3/D5-branes are relatively simple to state.  They consist of the 
following constraints on $H$ and $K$:
\eqr
0&=&\partial_y\partial_rK \nonumber \\
0&=&\partial_rH \label{constraints}\\
0&=&\partial_y(H^{-1}\partial_y(KH^{-1})) \nonumber \\
0&=&\partial_\rho (H^{-1}\partial_y(KH^{-1})). \nonumber \\
\rqe
The first two equations in the above are simple to solve.  However, the last
two are much less trivial when there is no smearing.  In fact it is not clear 
to me that there are non-smeared solutions to the above constraints.  
Finally there are source equations:
\eqr
\rho_{D3}&=& \frac{1}{r^2}\partial_r(r^2\partial_rK)
+\frac{1}{\rho^2}\partial_\rho(\rho^2\partial_\rho (H^{-1}K))
+\frac{1}{2}\partial_{y}^2(KH^{-1})^2 \\ 
\rho_{D5}&=&\frac{1}{\rho^2}\partial_\rho(\rho^2\partial_\rho H) + 
\partial_y^2K 
\nonumber
\rqe
  
There are two distinct sets of boundary conditions one can impose.  
The first set are 
appropriate for assymptotically flat backgrounds.  In this case, 
$K,H\rightarrow 1$ as $\rho\rightarrow\infty$.  
Another possibility is to
impose the condition that the geometry approach the AdS$_5\times$S$^5$
near-horizon geometry of the D3-branes as one moves away from the D5-branes
in either the $y$ or $\rho$ directions.

\subsection{Semi-localized solutions}
Although the exact equations for a localized solution of intersecting 
D3/D5-branes were presented above, they are non-linear and, at
least at the moment, 
seem intractable.  A considerable simplification occurs if we
look for solutions in which the D5-branes are smeared in the $y$ direction.
In this case one gains a new isometry with Killing vector $\partial_y$.  
The functions $K$ and $H$ are then independent of the $y$ coordinate. 
The constraints (\ref{constraints}) are satisfied if $H=H(\rho)$ and 
$K=K(\rho,r)$.  The source equations are as follows:
\eqr
\rho_{D5} &=& \frac{1}{\rho^2}\partial_\rho (\rho^2\partial_\rho H)\nonumber \\
\rho_{D3} &=& \frac{1}{r^2}\partial_r (r^2 \partial_rK)+
\frac{1}{\rho^2}\partial_\rho (\rho^2 \partial_\rho (H^{-1}K)) 
\rqe
Another simplification due to the smearing is that the NS-NS 3-form $H_3$ 
vanishes identically.  

Even though this system of equations is much simpler than the one for 
localized intersections it is not completely trivial.  Nevertheless, it
is possible to analyze the system in certain assymptotic regimes.  To
start with it is possible to find the exact solution to the first of the
two source equations:
\eq
H = 1 + \frac{q}{\rho}.
\qe
Although we are imposing boundary conditions appropriate to the zero slope 
Maldacena limit, the $1$ appearing in $H$ cannot be ignored since
it is not subdominant.  To see this let us try to determine what $q$ is.
Since $q$ counts the number of D5-branes, it must be proportional to
$n_5$, the number of D5-branes.  Secondly, we will consider smearing to be
a result of compactifying $y \sim y+2\pi R$, and then replacing the 
delta function in the source equation by the inverse volume of $y$.  Thus
$q\propto R^{-1}$.  Finally, we should have $q\propto n_5\alpha '/R$.  In
the Maldacena limit we keep $\rho /\alpha '$ fixed,
thus the $1$ does not become sundominant.  Also, as we shall see, to
get the appropriate AdS$_5\times$S$^5$ assymptotic geometry we must retain
the $1$.

We have not been able to find a closed expression for $K$
by solving the second equation.
Nevertheless, we can solve for $K$ in two different assymptotic regimes.
The first regime is when we are far away from the D5-branes.  This regime
is characterized by $q/\rho\ll 1$, thus $H\rightarrow 1$.  The second source
equation then becomes a six-dimensional flat Laplacian and the solution
for $K$ is simply:
\eq
K = \frac{R_0}{(r^2 + \rho^2)^2}.
\qe
In other words we have AdS$_5\times$S$^5$ geometry, which is just the
assymptotic geometry we wanted.  On the other hand we can also solve the 
equation in the regime in which we are very close to the D5-branes.  In 
this case $q/\rho \gg 1$, and hence $H\rightarrow q/\rho$.  In this case
we can re-write the source equation as:
\eq
\frac{1}{r^2}\partial_r (r^2 \partial_r(\frac{\rho K}{q}))+\frac{1}{q\rho}
\partial_\rho (\rho^2 \partial_\rho (\frac{\rho K}{q})) = 0.
\qe
In terms of $\eta = 2\sqrt{q\rho}$ the operator acting on $\rho K/q$ in the
above equation is:
\eq
\frac{1}{r^2}\partial_r (r^2 \partial_r)+\frac{1}{\eta^3}
\partial_\eta (\eta^3 \partial_\eta ).
\qe
This is just the sum of two radial Laplacians in $3$ and $4$ dimensions 
respectively.  A solution for $K$ which only depends on the $7$ dimensional
radius is given by:
\eq
K = \frac{qQ}{\rho (r^2+4q\rho)^5/2}+\frac{q}{\rho}
\qe
Where $Q$ is a constant.  This solution is valid in the $q/\rho \gg 1$ 
regime.  

Unfortunately we have been unable to make further progress with even the
simplified problem of smeared branes.  The problem as it stands is well-posed
and the differential equation is seperable.  Nevertheless new insights seem to
be needed to make progress.  We leave this to future work.

\section{Conclusions}
In this paper we developed a supersymmetric ansatz for general D3/D5-brane
webs.
We then focused on the subcase of infinite intersecting 
D3/D5-branes.  In that case we reduced the problem to a set of differential
equations involving two unknown functions.  These equations while simple 
to state are non-linear and we were unable to solve them exactly.  
We could however determine the
solution in certain asymptotic regimes in the smeared approximation.  

One of our main motivations for undertaking this project was to study AdS$_4$ 
branes in AdS$_5$ as suggested by \cite{kr}.  While we were not able to 
confirm the claims of \cite{kr} we have established the general structure
of the solution and posed the problem in a simple form.  
We hope that the results 
presented in this paper will contribute to a renewed effort to find the full
geometry of the D3/D5-brane system.  Some progress has already been made in
identifying certain features of the system in \cite{dfo,egk,lp,st}.

\section{Acknowledgements}
I would like to thank Cecilia Albertsson, Mario Caicedo,
Justin David, Ami Hanany, Tasneem Zehra Husain, Andreas Karch, 
Esko Keski-Vakkuri, 
Subir Mukhopadhyay, Kazutoshi Ohta, Markus Quandt, Martin Rocek, Martin 
Schvellinger, 
Douglas J. Smith and David Tong for valuable discussions. 
I would also like to thank the Helsinki Institute of Physics, the Institute 
for Advanced Study, and the Aspen Center for Physics for their hospitality
at various stages of this project.  I acknowledge support from the Swedish
Vetenskapsr{\aa}det and the Department of Energy under cooperative research
agreement\#DF-FC02-94ER40818.

\end{document}